# Solar Transients disturbing the Terrestrial Magnetic Environment at Higher Latitudes

Parvaiz A Khan[1], Sharad C Tripathi[1]*, O A Troshichev[2], Malik A Waheed[1], Aslam A M[1] and A K Gwal[1]

1. Space Science Laboratory, Department of Physics, Barkatullah University, Bhopal – 462026, M.P., India
2. Arctic and Antarctic Research Institute, St. Petersburg, 199397, Russia

***Email:** risingsharad@gmail.com

## Abstract

Geomagnetic field variations during five major Solar Energetic Particle (SEP) events of solar cycle 23 have been investigated in the present study. The SEP events of 01 October 2001, 04 November 2001, 22 November 2001, 21 April 2002 and 14 May 2005 have been selected to study the geomagnetic field variations at two high-latitude stations, Thule (77.5°N, 69.2°W) and Resolute Bay (74.4°E, 94.5°W) of the northern polar cap. We have used the GOES proton flux in seven different energy channels (0.8-4 MeV, 4-9 MeV, 9-15 MeV, 15-40 MeV, 40-80 MeV, 80-165 MeV, 165-500 MeV). All the proton events were associated with geoeffective or Earth directed CMEs that caused intense geomagnetic storms in response to geospace. We have taken high-latitude indices, AE and PC, under consideration and found fairly good correlation of these with the ground magnetic field records during the five proton events. The departures of the H component during the events were calculated from the quietest day of the month for each event and have been represented as $\Delta H_{THL}$ and $\Delta H_{RES}$ for Thule and Resolute Bay, respectively. The correspondence of spectral index, inferred from event integrated spectra, with ground magnetic signatures $\Delta H_{THL}$ and $\Delta H_{RES}$ along with Dst and PC indices have been brought out. From the correlation analysis we found very strong correlation to exist between the geomagnetic field variation (ΔHs) and high latitude indices AE and PC. To find the association of geomagnetic storm intensity with proton flux characteristics we derived the correspondence between the spectral indices and geomagnetic field variations (ΔHs) along with the Dst and AE index. We found a strong correlation (0.88) to exist between the spectral indices and ΔHs and also between spectral indices and AE and PC.

## Introduction:

During high energy solar transients, like Solar Flares, huge amount of energy is released from the Sun in very short duration of time. The release of energy is manifested either in the form of radiation across entire electromagnetic spectrum or mass ejections from the active regions commonly known as Solar Flares and Coronal Mass Ejections, respectively. Coronal Mass Ejections (CMEs) are large-scale expulsions of plasma from the Sun, driving coronal material into interplanetary space which can generate substantial disturbances in the ambient solar wind. The interplanetary counterparts of CMEs known as ICMEs are believed to be the prime cause of geomagnetic storms (Gosling et al., 1991; Bothmer et al., 1995; Webb et al., 2000; St. Cyr et al., 2000; Cane et al., 2000; Tsurutani et al., 2006). Every CME cannot lead to a geomagnetic storm, only those CME which are directed towards Earth or have a significant earthward velocity component, can cause geomagnetic storms. Therefore the occurrence of a front-side halo CME is also responsible for producing a geomagnetic storm and monitoring of these CMEs plays a central role in geomagnetic storm forecasting (Luhmann, 1997). However, even if an ICME reaches Earth, it may still not be highly geoeffective. An additional requirement is the Interplanetary Magnetic Field (IMF) having a southward component Bz of sufficient magnitude and duration (Bothmer et al., 1995). It has been noted that sometimes southward magnetic field is likely to occur within or ahead of most ICMEs (Webb et al., 2001).

There have been a number of studies investigating the possibility of CME speed as an indicator of CME geoeffectiveness. This is a concept often being used for space weather forecasting, where a 'fast' halo CME is considered as a high risk for adverse space weather, whereas, a 'slow' CME is interpreted as a smaller risk. There is a tendency of CME speed to some extent that it can be used as an indicator of CME geoeffectiveness (Srivastava and Venkatakrishnan, 2002; Kim et al., 2005), since some strongest magnetic storms were observed to be generated in the aftermath of fastest CMEs. Yurchyshyn et al. (2004) established similar relations between CME speeds, the magnitude of Bz, and the geomagnetic Dst index, while Gonzalez et al. (2004) showed that the expansion speeds of halo CMEs associated with magnetic clouds are related to the peak Dst index of the resulting magnetic storms. Several other studies including Gopalswamy et al. (2000) and Cane et al. (2000) have reported relationships between the CME's projected and travel speeds which, even though accompanied by a large scatter of the data, give some credence to the use of projected speeds as a proxy for the true travel speed of CMEs representing CME geoeffectiveness.

Another important property of the CME that can be used, to some extent, to indicate the geoeffectiveness of CME is the energetic particle flux associated with ICME. The CMEs moving with high speed accelerate the heliospheric particles present in the ambient solar wind to very high energies. Such high energetic particles are also referred as Solar Energetic Particles (SEP) and the abrupt enhancement of SEP flux is commonly known as solar energetic particle event. Forbush (1946) was first to observe the solar energetic particles during large solar flare events of 1942 as sharp enhancements in the intensity in the ion-chamber. The strong magnetic storms are often found to be accompanied by enhanced influx of energetic protons and ions. SEP events lasting for more than a day are thought to be generated primarily by strong shock waves propagating from the solar corona, through the interplanetary medium, to the geospace (Cane et al., 1987; Reames, 1999; Kahler, 2001). These particle events exhibit a close association with CMEs, and it is currently thought that the large, gradual, and long-lived SEP events are due to CME-driven shocks (Gosling, 1993; Reames, 1999). A weak correlation (broad scatter) between Peak SEP flux intensities and CME speed have been reported (Kahler et al., 1978; Cane et al., 1998) which suggests that SEP flux do not depend upon the CME speed but on some other factors, such as conditions preceding CMEs (Kahler and Vourlidas, 2005) or the presence of preexisting SEPs in the ambient medium (Kahler, 2001; Gopalswamy et al., 2004).

Particle signatures have been used as a tool to probe the large-scale magnetic structures of ICME and resulted in a close relationship between CMEs and ICMEs with energetic particles as they pass by the Earth (Richardson, 1997; Malandraki et al., 2005). From the fact that faster CMEs tend to generate stronger magnetic storms, and from the reported relations between CME speed and SEP flux intensities, we could expect some sort of relation between SEP fluxes and the strength of magnetic storms. SEPs observed at 1 AU convey useful information on the shocks driven by CMEs and it is not unreasonable to seek whether SEP flux characteristics somehow can be used to indicate CME geoeffectiveness. Smith et al. (2004) used observations of 47-65 keV ions, to predict the arrival of interplanetary shocks hours before they arrive at Earth. They suggested that the combination of halo CME observations and 47-65 keV ion enhancements exceeding a certain threshold can be used to improve predictions of magnetic storms. They emphasized that the discriminating features of SEP flux characteristics are important in the accessing the geoeffectiveness of CMEs in producing intense geomagnetic storms. In their study, Gleisner and Watermann (2005) discussed the role of SEP fluxes (≥10 MeV) as a tool to discriminate halo CMEs followed by strong magnetic storms from those not followed by strong storms, while Valtonen et al. (2005) demonstrated the feasibility of using 1-110 MeV protons to evaluate the geoeffectiveness of halo CMEs. Rawat et al., (2006) showed that SEP events with high flux levels or a plateau after the shock passage produce much more intense storms than the events where the SEP flux levels decrease after the shock passage even if the total flux values are similar. Gleisner and Watermann (2006) studied the 81 full disc halo CME for which alerts were issued by Naval Research Laboratory and concluded that SEP flux enhancements would be more efficient than CME speeds as an indicator of CME geoeffectiveness.

In the present study, we investigated the role of spectral index of the SEP events as an indicator of CME geoeffectiveness. We have taken five SEP events under consideration out of which four were associated with full halo CMEs and produced intense geomagnetic storms. We have found that during SEP events the high latitudes geomagnetic records show sharp depression from the normal value and at the same time the spectral index exhibits a strong correlation with the storm intensity index Dst as well as high latitude geomagnetic indices.

**Event Selection Criterion:**

During geomagnetic field variations at two high latitude stations Thule (77.5°N, 69.2°W) and Resolute Bay (74.4°E, 94.5°W) of the northern polar cap, we have selected five SEP events of 01 October 2001, 04 November 2001, 22 November 2001, 21 April 2002 and 14 May 2005 of the solar cycle 23 under the study which have flux value >$10^3$-$10^4$ pfu for the protons of energy >10 MeV which are under the S3 (Strong) NOAA Space Weather Scale for Solar radiation storms using NOAA definition of proton events having maximum flux of >10 particles/$cm^2$s.sr for >10 MeV protons, to understand the role of particle precipitation at higher geomagnetic latitudes. The various characteristics of these proton events along with the characteristics of associated Solar Flares and Coronal Mass Ejections are tabulated in table 1.

**Data Sets and Their Sources**

To study the geomagnetic field variations during the selected five SEP events, we have used data sets from both space and ground based systems. Solar flare characteristics were derived from the measurements of GOES satellite where as proton flux data were taken from the observations of GOES and ACE spacecrafts orbiting in geostationary orbit and at L1 point, respectively. The X-ray sensors onboard GOES continuously monitor the solar

X-ray flux output in two different pass bands (0.5 - 4 Å and 1 - 8 Å). The GOES mission also measures the solar proton flux in seven different energy bands ranging from 0.8 MeV-500 MeV. The Electron Proton Alpha Monitor (EPAM) instrument onboard ACE satellite also measures the proton flux at L1 point in eight different energy bands starting from 0.04 MeV to 4.75 MeV (Gold et al., 1998). These instruments provide the proton flux with different temporal resolution. However, for our study we have used the proton flux data with five minute time resolution. The Large Angle Solar Coronagraph (LASCO) on board SOHO spacecraft monitors the Coronal Mass Ejections released from the Sun. The various characteristics of CMEs like onset time, speed etc, required for study, have been taken from the LASCO observations (Brueckner et al., 1995). The Interplanetary Magnetic Field and solar wind data were taken from the MAG (Smith et al., 1998) and SWEPAM (McComas et al., 1998) instruments onboard ACE spacecraft, respectively. The state of magnetosphere is described by a number of magnetic activity indices derived specifically both for low and high latitudes. In our study we have used the polar cap magnetic activity index (PC) for northern polar cap (Thule) and auroral electrojet index (AE) with one minute resolution. The Dst index has been used for characterizing a geomagnetic storm. The data for geomagnetic indices have been taken from OMNI Data Web. The geomagnetic field component data were taken from National Geophysical Data Center (NGDC). We have used the one minute resolution data of the horizontal component of geomagnetic field for two high latitude stations Thule (77.5°N, 69.2°W) and Resolute Bay (74.4°E, 94.5°W). The departures of the horizontal component (H) of the geomagnetic field during the storms were calculated with reference to that of quietest day of the month for each event, by subtracting the storm day values from quietest day values, and have been represented as $\Delta H_{THL}$ and $\Delta H_{RES}$ for Thule and Resolute Bay, respectively.

**Analysis and Results:**

Solar cycle 23 witnessed a number of strong SEP events with peak particle flux reaching of the order of $10^4$ pfu. We have considered only five major particle events observed during the peak phase of the cycle 23. The GOES satellite observations showed that all these particle events were observed following either the X or M class solar flares. The case study for 04 November 2001 event has been taken as a representative of all the selected events and hence discussed herein, while other events are listed in the table 1.

**04 November 2001**

The most intense SEP event of solar cycle 23 was observed on the 04 Nov 2001 as reported by GOES and ACE missions. An intense solar flare of X1 class, produced by the active region 9684 was recorded by GOES on 04 Nov 2001 with start time 16:03 UT. After 45 minutes of the flare maximum at 16:20 UT a proton shower started which was recorded by GOES and ACE particle detectors. The proton event started on 04 Nov 2001 at 17:05 UT and achieved the peak flux value of 31,700 pfu on 06 Nov 2001 at 02:15 UT. The time profile of X-ray flux and proton flux recorded by both GOES and ACE spacecrafts, in different energy channels covering both low and high energy range, are depicted in Figure 1. A halo CME travelling with a speed of 1810 km/s was associated with the event as detected by the LASCO at 16:35 UT on the same day. The fast moving CME drove an interplanetary shock which was noticed by the abrupt increase in the particle flux. After the shock the flux remained high for long time, a plateau of flux, as described by Rawat et al. (2006). The Interplanetary Magnetic Field components, Solar Wind conditions and Geomagnetic Field variations during the 05 Nov to 09 Nov 2001 are shown in Figure 2 and 3. After the shock the steady and smooth variation in IMF were replaced by rapid fluctuation of Bz component, turning southward achieving peak value of -68 nT at 01:48 UT on 06 Nov 2001. The interaction of the shock with the magnetosphere resulted in an intense geomagnetic storm with minimum Dst -292 nT on 06 Nov 2001. The dynamic interaction was reflected by the significant increase of the AE and PC indices. A remarkable decrease of $\Delta H_{THL}$ and $\Delta H_{RES}$ can be clearly noticed from the Figure 2. IMF Bz made a second turn and in accordance $\Delta H_{THL}$ and $\Delta H_{RES}$ also showed a slight decrease again before the complete recovery of $\Delta H_{THL}$ and $\Delta H_{RES}$.

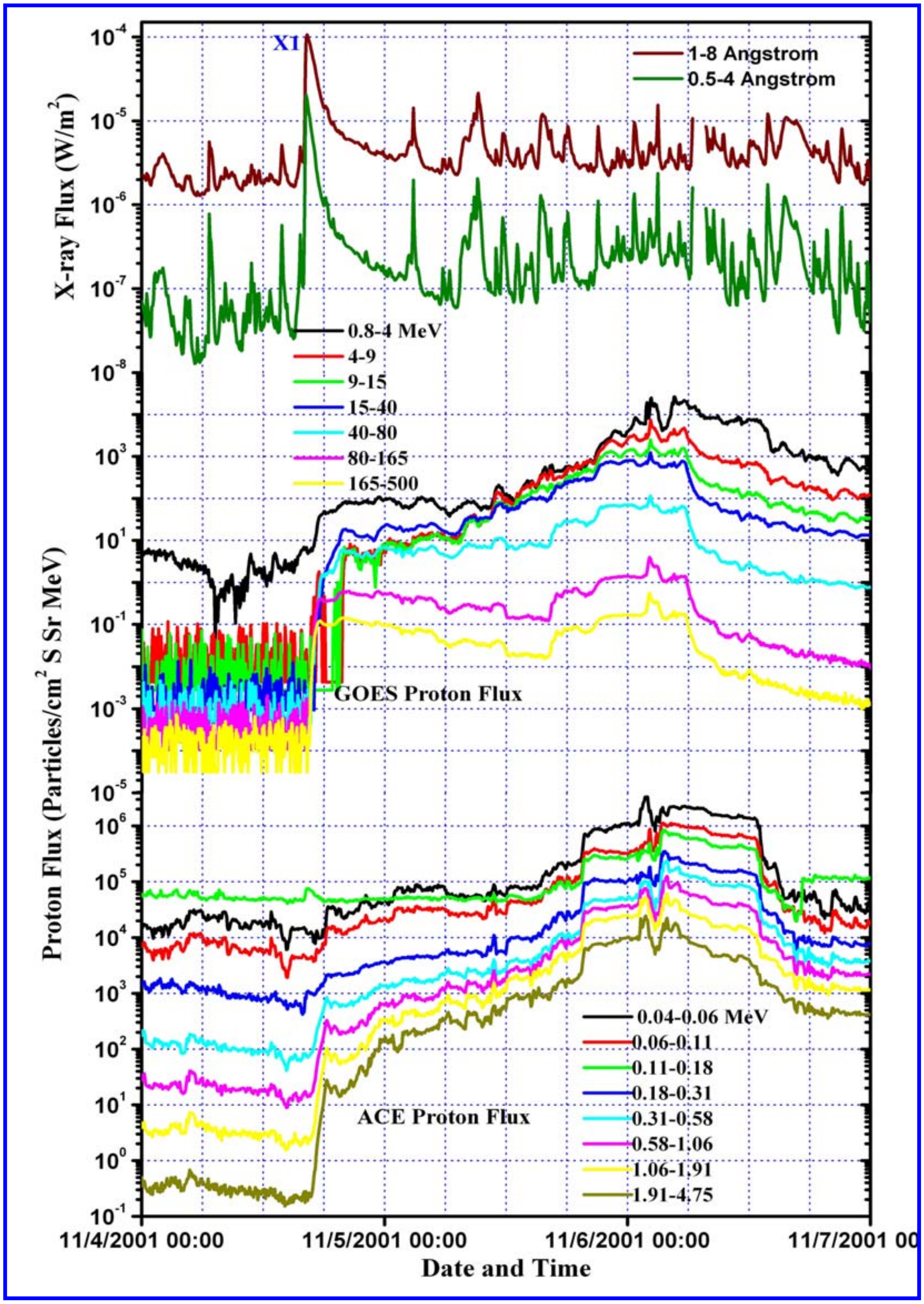

**Figure 1**: Time profile of Solar X-ray radiation with GOES and ACE protons observed on 04 November 2001.

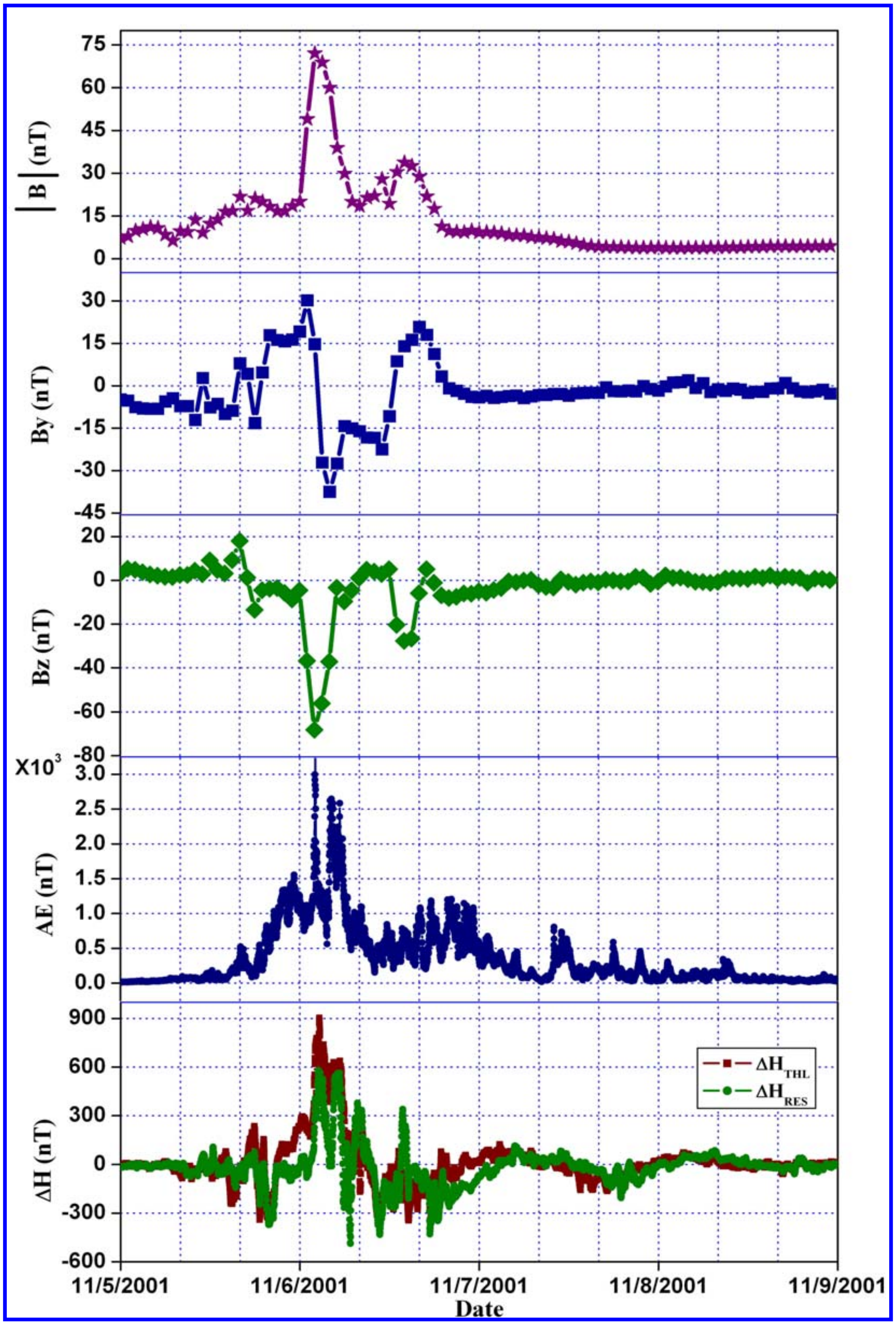

1 **Figure 2**: Temporal variation of IMF components along with variation of auroral electrojet
2 index and H component of ground magnetic field during 05-09 Nov 2001.

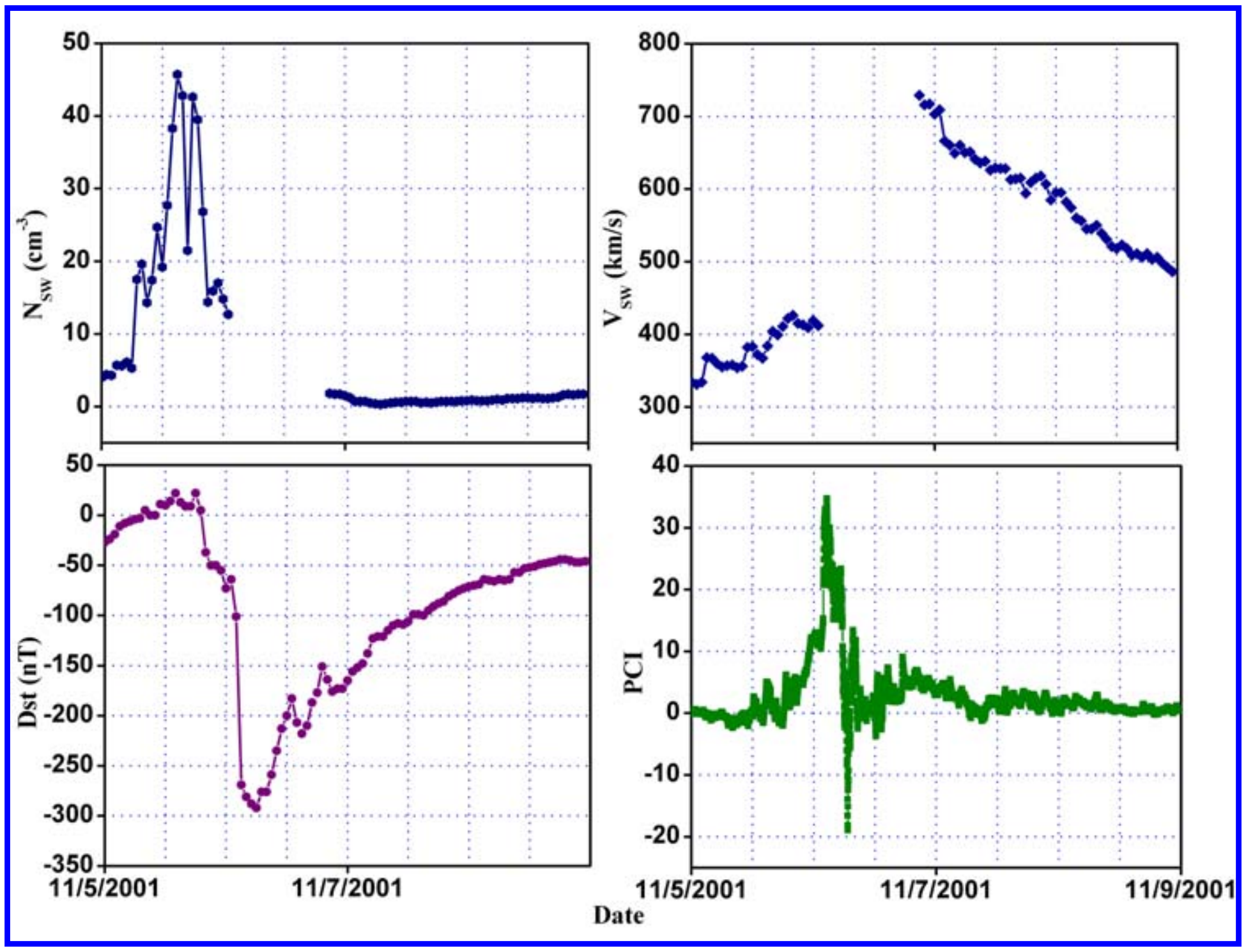

**Figure 3**: Temporal evolution of Dst with varying Solar Wind parameters and Polar Cap Index during 05-09 Nov 2001.

Dynamic interaction between solar wind and magnetosphere is reflected both in high and low latitude geomagnetic field records. To bring out the correspondence between the solar influences and their geomagnetic signatures at high latitudes, we have taken the AE and PC index and departures (from quiet time) of H component of geomagnetic field (ΔH) at two high latitude stations Thule and Resolute Bay. During solar wind-magnetosphere interaction the energy carried by the solar wind is thought to be deposited into the magnetosphere through dominant mechanism of magnetic reconnection (Phan et al., 2000). High latitude magnetic field perturbations due to this energy transfer are defined by an index known as auroral electrojet AE index (Koskinen and Tanskanen, 2002). The interaction of energetic charged particles with ambient electric and magnetic fields generates various current systems like DP1 and DP2. The magnetic disturbances due to ionospheric and field aligned currents, *viz.* Hall current at high latitudes, are described by an index known as polar cap index (PC) (Troshichev et al. 1988).

In order to do the correlative study of polar cap index (PC), auroral electrojet index (AE) with ground magnetic field signatures we have chosen a specific time interval which is the time between start and complete recovery of the depression in the H component ($\Delta H$). The results of correlation analysis are presented in figure 4 for the 04 Nov 2001 event. We have found the similar type of results for all the events under consideration to exist a linear fit for the scatter of AE and PC index. A very strong correlation with correlation coefficients of 0.72-0.89 exits between the two indices. The correlation coefficients are shown in the table 2 for all the selected events and magnitude of correlation is found to be independent of the intensity of the storm, particle or CME characteristics. Even in the case of weak storms the strong correlation is maintained *viz.* for the geomagnetic storm produced by the 01 Oct 2001 solar transients the minimum Dst was only -57 nT but the correlation coefficient between AE and PC index is found to be 0.84. Whenever there are geomagnetic disturbances the high latitude indices AE and PC exhibit a strong correlation, which are in accordance with the results of Takalo and Tinomen 1998, irrespective of storm or solar transient characteristics. It is believed that PC can be used as a first available indicator of DP2 and DP1 activity in the Polar Regions. Compared to AE, PC has the limitation that during summer it can only be used for weak or southward Bz due to the effect of the so-called "reversed convection events."

The correlation between the high latitude geomagnetic field variations ($\Delta H_{THL}$ and $\Delta H_{RES}$) and auroral electrojet AE index, for 04 Nov 2001 event, is shown in upper panel of Figure 4. A second order fit with good correlation exists for all the five events for both the stations. The correlation coefficient between AE and $\Delta H_{THL}$, $\Delta H_{RES}$ ranges from 0.50 to 0.85 depicting a very good agreement between the changes in AE index and ground magnetic field perturbation during magnetically active conditions. The magnitude of correlation is independent of the storm intensity *viz.* for moderate storm of 21 Apr 2002 (Dst -57 nT) the correlation coefficients between $\Delta H_{THL}$ and AE and $\Delta H_{RES}$ and AE are 0 .85 and 0.64, respectively, while for intense storm of 4 Nov 2001 (Dst -292 nT) the coefficients are 0.72 and 0.58. Similarly for 22 Nov 2001 (Dst -221 nT) the coefficients are 0.66 and 0.84 for Thule and Resolute Bay, respectively. Therefore ground magnetic field perturbations at high latitudes do not seem to have association with the intensity of the storm or associated solar events. Irrespective of the characteristics of a solar event or storm, the high latitude geomagnetic field variations follow a very good association with AE during the event. Therefore AE is found to be a very effective index to represent the geomagnetic conditions at high latitudes.

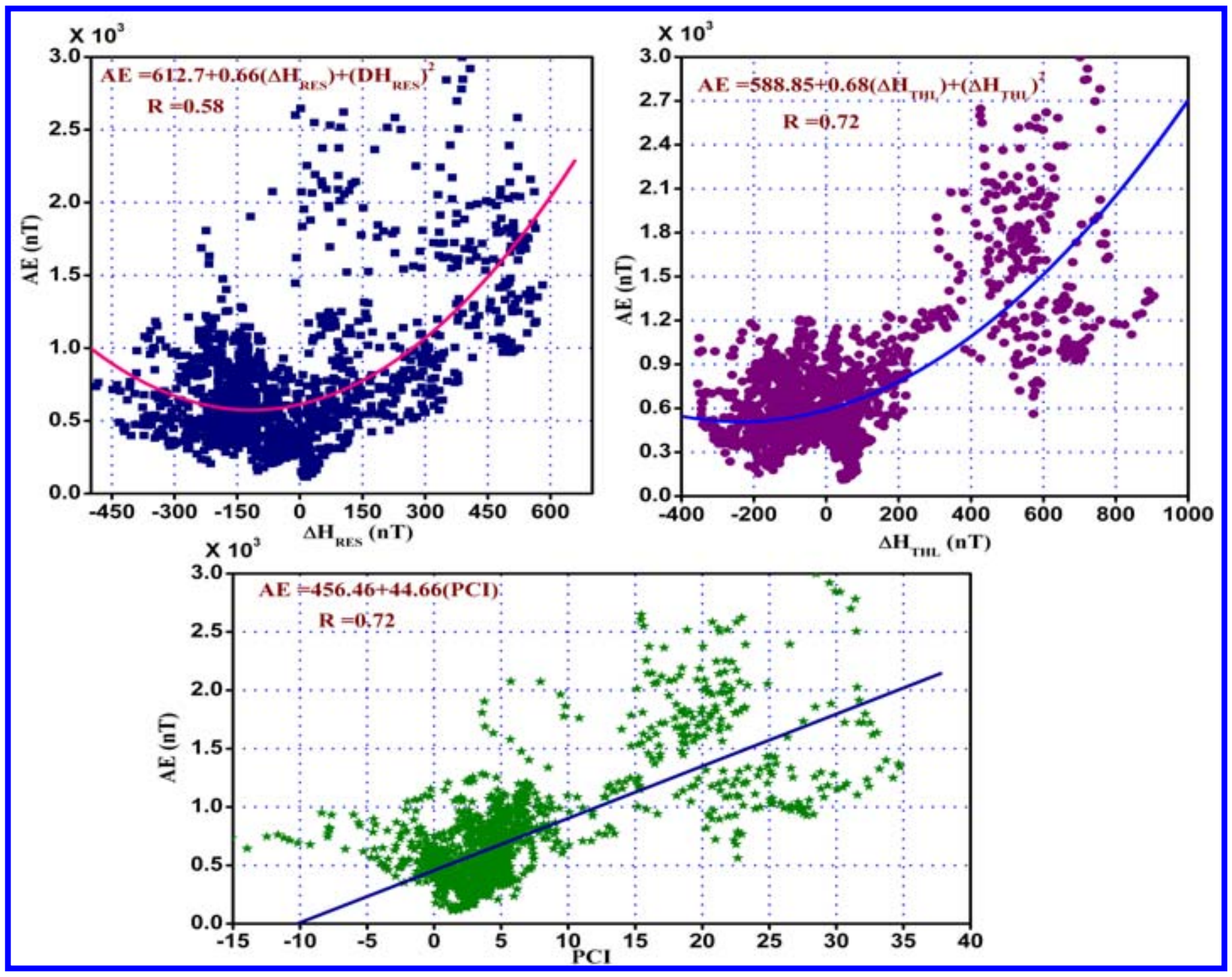

**Figure 4**: Scatter of AE index with the ground magnetic field perturbations and polar cap index for the 04 November 2001 event.

Event integrated spectra has been constructed for each particle event and spectral index has been computed by fitting them using power law. The correspondence of spectral index with Dst and PC indices is shown in Figure 5 and 6. A linear fit exits between the AE index and proton spectral index as well as between PC index and proton spectral index with correlation coefficients of 0.88 and 0.89, respectively, indicating a strong correlation between these indices. Hence, spectral index can be treated as an effective index to associate the flux characteristics of a particle event with the geoffectiveness of a halo CME in producing geomagnetic storms which is in accordance with the findings of Tripathi et al., 2013 that both the particle events and CMEs are associated with each other having the same process of genesis. A number of studies earlier have already highlighted the role of flux characteristics of particle events to access the geoffectiveness of a CME (Smith et al., 2004, Valtonen et al., 2005, Gleisner and Watermann, 2006). Our analysis also shows that the flux characteristics of a particle event associated with a CME seems to have a property of indicating the goeffectiveness of a halo CME. The index which can better reflect it is the spectral index of the particle event. There is a fundamental consensus that gradual SEP events result from acceleration of particles at shocks, where the strength and spatial structure of the shock determines the characteristics of the observed SEP fluxes. At the same time, shock compression of magnetic fields is an important source of the large-magnitude IMFs that is a key factor in the generation of magnetic storms (Jurac et al., 2002). Preceding CMEs, preexisting energetic particles, and the density structure of the extended corona all have an influence on the production of SEPs, being a common consensus on energetic particle production, a better understanding of the interrelations amongst these factors, and a more complete mapping of their relations with geomagnetic storms of different strengths, might allow us to use near-Earth observation of SEPs to improve geomagnetic forecasting.

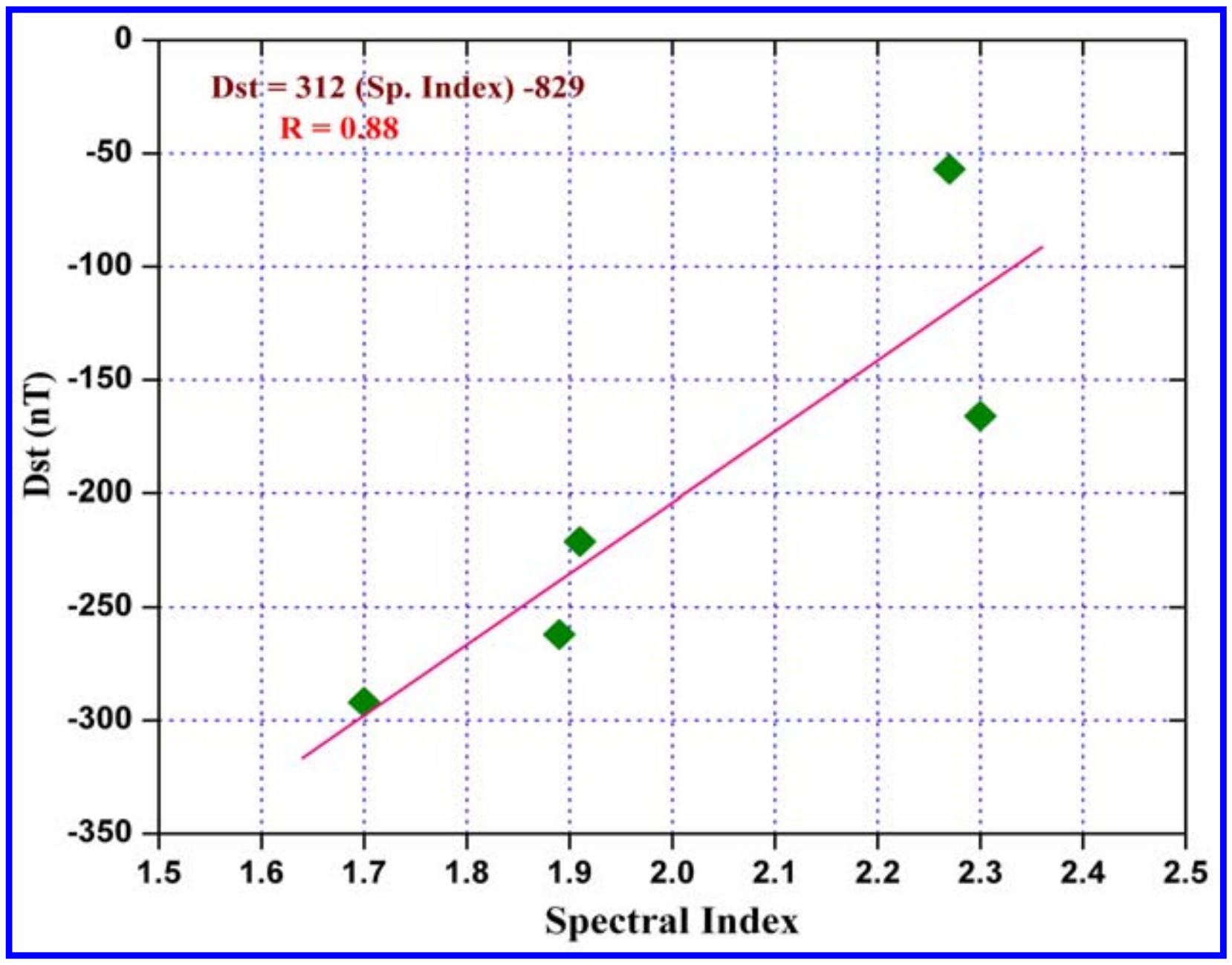

**Figure 5**: Correlation between the spectral indices and Dst for all the five events.

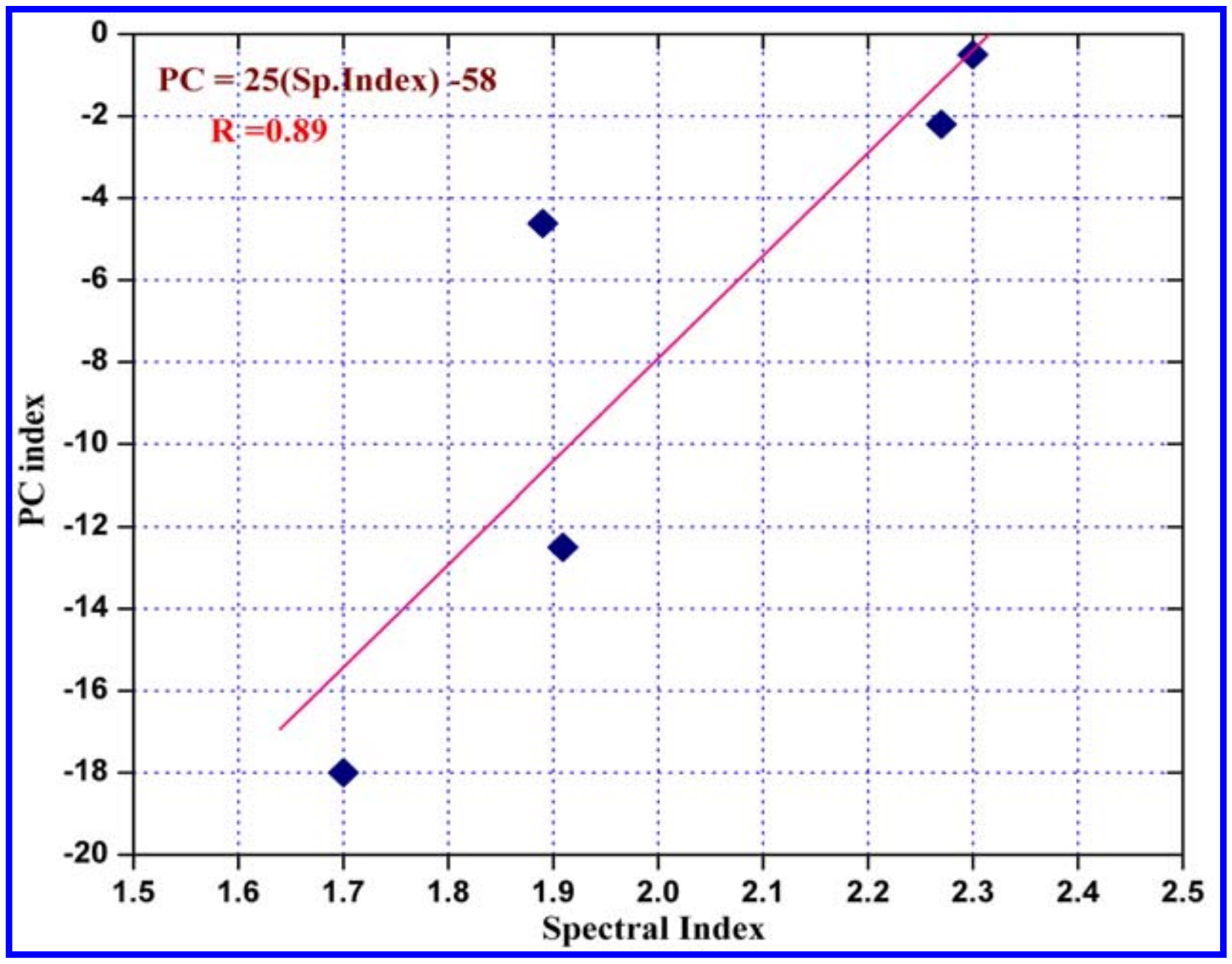

**Figure 6**: Correlation between the spectral indices and PC index for all the five events.

**Conclusion:**

The high latitude geomagnetic field is found to be affected strongly by the solar disturbances. During geomagnetic storms the smooth variation of horizontal component of high latitude geomagnetic field undergoes periods of large decrease as well as fluctuates rapidly. The high latitude geomagnetic field perturbations ($\Delta H_{THL}$ and $\Delta H_{RES}$) have strong correlation with AE index during event time and a strong correlation is found not only during intense storms but also during weak storms. The polar cap index (PC) correlates well with the AE index during event time irrespective of the storm intensity or particle flux characteristics. Instead of CME speed, SEP flux characteristics can be used as an effective indicator of CME geo-effectiveness as SEPs are thought to be produced from solar mass motions loaded in CMEs and accelerated at the shock front. The spectral index of the particle event follows an extremely strong correlation with the Dst as well as PC index indicating that SEP flux characteristics can be an important index to show the geoeffectiveness of CME in producing geomagnetic storms along with the CME speed.

**Acknowledgements:**

Authors (Sharad C. Tripathi and Parvaiz A. Khan) are thankful to University Grant Commission, New Delhi for the financial support under Basic Scientific Research (BSR) Fellowship scheme during the work. Special thanks go to Dr. O A Troshichev for his meticulous review and criticism of the content during Mr. Tripathi's visit to Addis Ababa, Ethiopia at AGU's Chapman Conference in November 2012 and for being a coauthor of the paper as well.

**Table 1:** SEP, CME and flare characteristics of five events along with minimum Dst, PCI and spectral index.

| SN | SEP Characteristics | | | CME | | | FLARE | | | | | Dst (min) | PCI | SP Index |
|---|---|---|---|---|---|---|---|---|---|---|---|---|---|---|
| | Date | Start Time (UT) | Peak Flux (pfu) | Onset Date | Time (UT) | Type | Date | Start Time (UT) | Class | AR | Location | | | |
| 1 | 1-10-2001 | 11:45 | 2,360 | 1-10-2001 | 05:30 | SW | 1-10-2001 | 04:41 | M9.0 | 9628 | S22W91 | -160 | -0.5 | 2.30 |
| 2 | 4-11-2001 | 17:05 | 31,700 | 4-11-2001 | 16:35 | Halo | 4-11-2001 | 16:03 | X1.0 | 9684 | N06W18 | -292 | -18.0 | 1.70 |
| 3 | 22-11-2001 | 23:20 | 18,900 | 22-11-2001 | 23:30 | Halo | 22-11-2001 | 22:32 | M9.0 | 9704 | S15W34 | -221 | -12.5 | 1.91 |
| 4 | 21-4-2002 | 02:25 | 2,520 | 21-4-2002 | 01:27 | Halo | 21-4-2002 | 01:51 | X1.0 | 9906 | S14W08 | -57 | -2.2 | 2.27 |
| 5 | 14-5-2005 | 05:25 | 3,140 | 13-5-2005 | 17:22 | Halo | 13-5-2005 | 17:10 | M8.0 | 10759 | N12E11 | -262 | -4.62 | 1.89 |

Table 2: Correlation Coefficients between AE and PC for all five events under consideration.

| Sr.No. | Date | Correlation Coefficient between AE and PC |
| --- | --- | --- |
| 1 | 01 October 2001 | 0.89 |
| 2. | 04 November 2001 | 0.72 |
| 3. | 22 November 2001 | 0.79 |
| 4. | 21 April 2002 | 0.84 |
| 5. | 14 May 2005 | 0.72 |